# Extension of the INFN Tier-1 on a HPC system

*Tommaso* Boccali[1], *Stefano* Dal Pra[2], *Daniele* Spiga[3], *Diego* Ciangottini[3], *Stefano* Zani[2], *Concezio* Bozzi[4], *Alessandro* De Salvo[5], *Andrea* Valassi[6], *Francesco* Noferini[7], *Luca* dell'Agnello[2], *Federico* Stagni[6], *Alessandra* Doria[8], *Daniele* Bonacorsi[9]

[1]INFN Sezione di Pisa, L.go B. Pontecorvo 3, 56127 Pisa (ITALY)
[2]INFN CNAF, Viale Carlo Berti Pichat, 6/2, 40127 Bologna (ITALY)
[3]INFN Sezione di Perugia, Via Alessandro Pascoli 23c, 06123 Perugia (ITALY)
[4]INFN Sezione di Ferrara, Via Saragat 1, 44122 Ferrara (ITALY)
[5]INFN Sezione di Roma1, P.le Aldo Moro 2, 00185 Roma (ITALY)
[6]CERN, Esplanade des Particules 1, P.O. Box 1211, Geneva 23 (SWITZERLAND)
[7]INFN Sezione di Bologna, Viale Carlo Berti Pichat 6, 40127 Bologna (ITALY)
[8]INFN Sezione di Napoli, Via Cintia, 80126 Napoli (ITALY)
[9]Dipartimento di Fisica e Astronomia, Viale Berti Pichat 6/2, 40127 Bologna (ITALY)

**Abstract** The INFN Tier-1 located at CNAF in Bologna (Italy) is a center of the WLCG e-Infrastructure, supporting the 4 major LHC collaborations and more than 30 other INFN-related experiments.
After multiple tests towards elastic expansion of CNAF compute power via Cloud resources (provided by Azure, Aruba and in the framework of the HNSciCloud project), and building on the experience gained with the production quality extension of the Tier-1 farm on remote owned sites, the CNAF team, in collaboration with experts from the ALICE, ATLAS, CMS, and LHCb experiments, has been working to put in production a solution of an integrated HTC+HPC system with the PRACE CINECA center, located nearby Bologna. Such extension will be implemented on the Marconi A2 partition, equipped with Intel Knights Landing (KNL) processors. A number of technical challenges were faced and solved in order to successfully run on low RAM nodes, as well as to overcome the closed environment (network, access, software distribution, … ) that HPC systems deploy with respect to standard GRID sites. We show preliminary results from a large scale integration effort, using resources secured via the successful PRACE grant N. 2018194658, for 30 million KNL core hours.

## 1 Introduction

Italian physicists have historically been major players in the design, construction and operations of the LHC detectors, via the funding of the Istituto Nazionale di Fisica Nucleare (INFN). In particular, Italy supports LHC distributed computing infrastructure via 10 WLCG[1] facilities, with a Tier-1 site in Bologna (Italy), at INFN-CNAF; currently, it provides about 10% of the WLCG Tier-1 total resources, with the exact share depending on the experiment. On top of that, CNAF supports non LHC computing activities linked to INFN research, with computing and storage provided to more than 30 experiments.
In the next decade the computing needs of INFN experiments, in particular those in the field of High Energy Physics (HEP), are projected to increase faster than the technology can provide at fixed cost during the same period [2]. Alternative directions are explored in order to overcome the resource problem; one particularly attractive requires the utilization of the High Performance Computing (HPC) facilities worldwide, expected to be at the Exascale (1 Exaflops = $10^{18}$ floating point operations per second) in the same time period.

This report describes a successful integration attempt between INFN-CNAF and an HPC system at CINECA (also in Bologna), underlying the challenges and the solutions deployed. The paper is organized as follows. Section 2 gives an overview of the INFN-CNAF computing center. Section 3 describes the Marconi A2 HPC systema at CINECA, where this work was performed. Section 4 reviews the general challenges faced by the LHC experiments to integrate their workflows on HPCs and more specifically on Marconi A2. Section 5 gives details about the work performed in each of the four experiments, in the software and computing areas. Results are given in section 6 and conclusions in Section 7.

## 2 INFN-CNAF center in Bologna

INFN-CNAF, a WLCG Tier-1 site, deploys "GRID-like" resources for a total of 30k CPU cores, 38 PB of disk and 90 PB of tape storage; it supports more than 30 research activities funded by INFN.

The facilities have been designed and deployed in order to satisfy the data intensive workflows from LHC experiments, and deploys computing nodes with high bandwidth connection to the local and remote storages. The center is connected via GARR/Geant to peers worldwide, with a total bandwidth of 200 Gbit/s; an additional link of 20 Gbit/s is provided for the general Internet. The single computing nodes are equipped with multicore x86_64 CPUs, with at least 3 GB / thread memory, and rotating / memory local disks of the order of 10 GB/thread. Outgoing connectivity from the single node is permitted without limitations, as needed by the workflows processing remote data via streaming protocols. Since 2018, a significant fraction (45%) of the CPU power has physically been deployed at CINECA (see next section), but it is logically seen as part of CNAF farm with the help of a dedicated connection realized via a pair of Infinera DCI, currently limited at 400 Gbps.

## 3 Marconi A2 system at CINECA

CINECA, a PRACE Tier-0 facility, currently hosts a system, Marconi [3], ranked no. 19 in the top500.org [4]. The Marconi A2 partition, of interest in the present study, deploys 3600 nodes equipped with 1 Xeon Phi 7250 (KNL) at 1.4 GHz, with 96 GB of on board RAM. Each CPU has 68 physical threads, with 4-way hyperthreading. The total 244,800 cores are rated at 11 PFlops as peak performance. In the standard configuration, the nodes do not have user access to on board disk but are connected via Omni-Path [5] to a large GPFS storage, and do not allow routing to IPs external to CINECA.

Resources at CINECA are partially provided within PRACE [6], via a call system which grants CPU-hours after a review process. The Italian LHC Community successfully applied to the "18th PRACE Project Access Call for Proposals", and was assigned a grant of 30 Million CPU hours on the Marconi A2 partition. The grant was requested for specific physics studies, but also for initiating an handshaking procedure with CINECA in view of future larger utilization of CINECA's systems for HEP computing.

Indeed, the INFN-CNAF Tier-1 and the next CINECA HPC (Leonardo, a pre-exascale system expected by 2021) will be partially co-located in a new facility, and any expertise acquired in running LHC workflows in the current system would be precious in planning the Leonardo infrastructure.

# 4 How to match a HPC system with LHC workflows

Matching LHC workflows with HPC systems is not a trivial effort. HPC facilities usually have strict site policies for security reasons and may be based on a variety of specialized hardware setups, so that they are characterized by:
- limited / absent external connectivity,
- user policies not allowing delegation of authentication,
- a range of architectures for CPUs (x86, ARM, Power9), GPUs or FPGAs,
- hardware setups which may include low RAM or no local disks,
- ad-hoc operating systems.

On the other hand, LHC workloads are designed to run on "standard WLCG sites", which at least in the first LHC phase had:
- uniform base architectures (x86 CPUs) and operating systems
- abundant RAM and local disk,
- full access to remote services like CVMFS for software installation
- the capability to use user-level virtualization,
- a common framework for accounting, traceability, with authentication and authorization delegated to external service.

As a consequence, integrating LHC workflows on HPC centers poses two rather distinct types of challenges, in the software area (e.g. efficiently exploiting low-memory many-core CPUs, or porting applications to GPUs) and in the computing area (e.g. managing job scheduling and software/data distribution). It should also be noted that HPC centers provide very high-speed inter-node connectivity, but most (not all) types of LHC workflows submit independent jobs on the different individual nodes, i.e. they use an HPC as a large cluster of nodes. These and other issues of the WLCG strategy towards HPC have been clarified with a white paper document [7], complemented with some more technical documents as in [8].

Given the unicity of HPC systems, solutions have to be found case by case, and are in general expensive in terms of operations and manpower. In the specific case under study, Table 1 shows the mismatch of configurations between a Marconi A2 computing node and a typical WLCG node; the color codes are according to the "severity" of the limitation as detailed in [8].

**Table 1.** Differences between a WLCG node and a standard Marconi A2 computing node.

| A typical standard Marconi A2 node is configured with | A typical WLCG node has |
| --- | --- |
| A KNL CPU: 68 or 272(HT) cores, x86_64, rated at ~¼ the HS06 of a typical Xeon | 1/2 Xeon-level x86_64 CPUs: typically 32-64 cores, O(10 HS06/thread) with HT on |
| 96 GB RAM, with ~10 to be reserved for the OS: only 0.3 GB/thread if all 272 HT threads are used | 2GB/thread, even if setups with 3 or 4 are more and more typical |
| No external connectivity | Full outbound external connectivity, with remote sw accessed via CVMFS mounts |
| No local disk (large scratch areas via GPFS/Omnipath) | O(20) GB/thread local scratch space |
| Access to batch nodes via SLURM; Only Whole nodes can be provisioned, with 24 h lease time | Access via a CE. Single thread and 8 thread slots are the most typical; 48+ hours lease time |
| Access granted to individuals | Access via pilots and late binding; VOMS AAI for end-user access |

In order to overcome the technical difficulties, several meetings were organized between LHC experiments' computing experts and CINECA management and sysadmins, where solutions were discussed, deployed and eventually tested.

In particular, the Marconi A2 node configuration was updated with:
- CVMFS mounts, which turned out to be useful also outside the HEP community, and were already considered by CINECA. Squids in order to lower the external traffic have been deployed by CINECA on edge nodes;
- external outbound networking was partially opened, with routing active to CERN and CNAF IP ranges;
- the singularity [9] virtualization tool was audited by CINECA, and green-lighted for deployment;
- a HTCondorCE was allowed on a CINECA edge node, with access to the external IP ranges as above, and the ability to submit to the internal SLURM[10] batch system connected to Marconi A2;
- (still to be commissioned) in order not to overload CINECA's GPN, a partial 40 Gbit/s connection was established on the Infinera private link.

In such conditions, all the experiments were able to prove basic processing functionalities, as necessary for the Grant application.

Pilots reach the HTCondorCE [11] via its public interface, and are submitted to the KNL nodes via SLURM. They call back the Experiments Workload Management Systems and receive payloads, which are executed inside a singularity sandbox, with a OS image available via CVMFS.

The software and calibrations needed to execute each payload are accessed via CVMFS, without an explicit local installation at CINECA.

Input and Output data follows different paths. Input data (if needed) are accessed directly using the Xrootd protocol if on CNAF storage; for other sources, not reachable due to the routing limitations, an Xrootd proxy has been deployed at CNAF to buffer the input files and then serve the executing processes. This was done in collaboration with the XDC EU Project [12]. Effectively, all remote files were accessible by the KNL nodes via the cache or directly. Output files were shipped to the CNAF storage system via SRM/Xrootd protocols, and as such registered in the Experiments' Data Management Systems.

With this strategy, the payloads can effectively access all the files in the LHC data federations, albeit with low bandwidth. In order to be able to compute at high CPU efficiency, low IO workloads must be carefully selected, via means specific to the single Experiments.

## 5 Workflows of LHC experiments on Marconi A2

### 5.1 ALICE

Usually, ALICE production pilot jobs are submitted to the GRID using an automatic procedure running on a dedicated user interface (vobox) installed in each site. The ALICE access to Marconi A2 resources is realized with a submission procedure customized for this test and based on HTCondor. In this specific case one of the CNAF voboxes, configured for HTCondor submission, is used to perform a manual submission of simulation jobs with the new O2 software framework [12] developed for ALICE Run-3. There are two reasons why we are not using production jobs:
1. the O2 software is not yet used in the current GRID workflow;
2. some specific configurations, not yet implemented, are required to run with full external connectivity.

It is worth noting that the result of this test is very relevant not only for testing HPC systems but also for the validation of O2 simulations.

The main features of the submitted jobs are:
- they are dispatched to WNs each one providing 68 physical cores (hyperthreading x4) and 86 GB RAM;
- input and output are read/written on CNAF ALICE storage via Xrootd;
- ALICE O2 software is available via CVMFS and run with singularity;
- simulations are performed with pythia8 with Pb-Pb collisions at 5.5 TeV per nucleon pair and using GEANT4 [14] as particle propagator in the detector material.

The use of Pb-Pb simulations allows to test all the functionalities of the O2 software and, in particular, the parallel propagation of particles in the same event with multi-thread processes (to reduce RAM usage).

In order to guarantee a reasonable time per job (within 12 hours), we limited the number of simulated events to 160. Simulation workflows are performed via different configurations, by varying the number of simultaneous independent processes (instances) and the number of parallel threads per instance (workers). Performances are characterized using metrics collected during the execution of each job by a dedicated script: wall time, output size, RAM usage peak.

The metrics collected so far show very good scalability of the simulation jobs up to a number of threads of the order of physical cores, O(68). Indeed, the number of simulated events per second increases linearly with the number of threads. Above that threshold, hyper threading kicks in, and we didn't observe any significant gain (see Figure 1). This may be related to the nature of jobs which are intrinsically high CPU consuming.

The splitting of threads over many independent instances/jobs doesn't change the performances, the time to simulate one event is depending only on the overall number of effective threads used. However, the usage of memory is sensitive to the number of independent jobs running on the machine. As expected the RAM required by the jobs increases both with the number of overall threads and the number of instances used. For example, a simulation (160 Pb-Pb events) performed with 64 workers/threads requires about 15 GB if the threads belong to a single job, or about 45 GB if threads are splitted in 16 independent jobs (4 threads each).

The actual conclusion at the end of this first exercise is the demonstration that ALICE is able to run Run-3 simulation jobs in a HPC system if well defined requirements are satisfied, as provided in our case both from the CINECA and CNAF staffs.

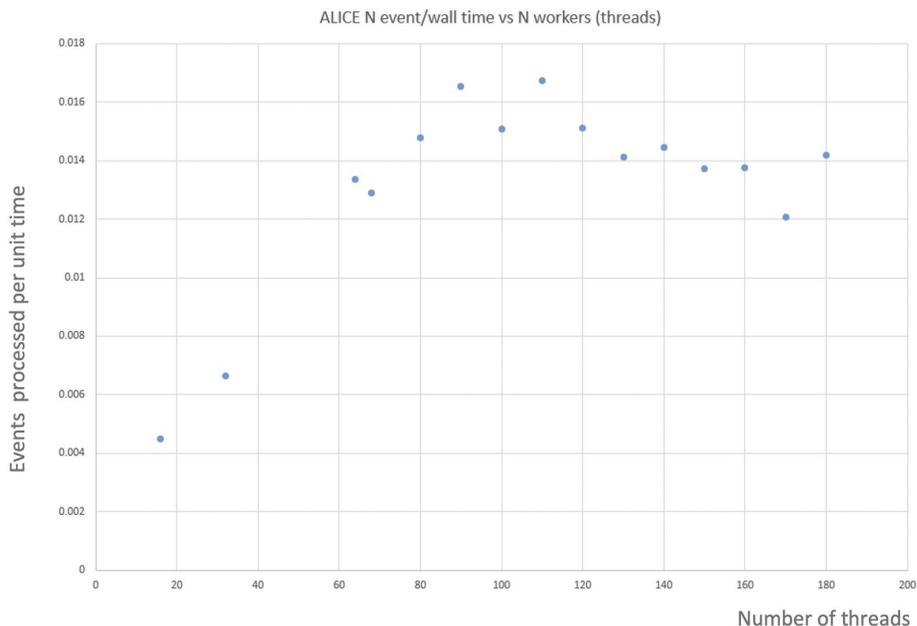

**Fig. 1.** Performance (as events produced per unit time) of a KNL node as a function of the number of threads..

## 5.2 ATLAS

The ATLAS Experiment uses the Production and Distributed Analysis (PanDA) system [15] to manage production and analysis tasks. The system is designed to fulfil the requirements of a data-driven workload management system, capable of operating at LHC data processing scale. PanDA is able to submit jobs to the sites, thus to the underlying resources, in a fully transparent way for the users, so is able to connect to different types of resources, including Grid, HPC and Cloud sites.

The CINECA HPC resources are seen by ATLAS as a standard HPC queue defined in PanDA, thus the central system sends pilots directly to the HTCondor Computing Elements (CE) available in the facility. The parameters of the queues, needed to identify and shape the task requirements, are defined in the ATLAS Grid Information System (AGIS) [16].

Once the pilots are running in the CINECA nodes, they contact the central PanDA Servers to gather the definitions of the jobs assigned to the specific queues, and execute the payloads.

Since the nodes in CINECA have 96 GB of RAM (with 86 usable by user processes), in order to avoid out-of-memory conditions, following the requirements of the software, only multicore jobs with 48 threads are allowed, and only simulations jobs are accepted by the queues, in order to avoid waste of resources or inefficiencies. The definition of the tasks that are allowed to run on the CINECA queues is also defined in AGIS.

ATLAS is using a model where both Multi-Processing (fork) and Multi-Threading (threads) are allowed, but the fully threaded option is still experimental, so it was not yet used in CINECA. The situation on the amount of used memory, so the maximum number of threads in use at the same time, will improve when the fully threaded infrastructure will be completed and validated.

Tests showed that ATLAS is actually able to run up to 128 processes in the current configuration, with low-memory simulation tasks, although in such cases the efficiency drops to ~80%, while with 48 threads the efficiency has been measured to be almost the same with respect to standard jobs running on ordinary Intel or AMD processors. A typical evolution with the multiprocess size is shown in Figure 2 (left).

**5.3 CMS**

The CMS Experiment uses a Workload Management System based on HTCondor, with workload submission divided into two steps.

As a first step, a GlideinWMS Factory [17] submits "pilots" to the Computing Elements (CE) at the facilities. These are converted into batch jobs with the site-specific technology (SLURM at CINECA), and are started according to site policies and resources shares.

The pilots, once executed, call back the CMS Global Pool [18], and register as available resources. In the case under study, the Marconi A2 nodes were registered as standard CNAF nodes, and workload were proposed matching generic CNAF requirements.

An R&D within CMS allowed to fine tune the matching system, via a cherry-picking method. The Marconi resources, when joining the CMS Global Pool, specify additional requests with respect to CNAF nodes, and in particular:
- ask for low memory jobs, since the standard RAM/thread is lower on KNL than on Xeon nodes at CNAF;
- ask for workload types ("Subtask names") known to need low IO (simulation and generation);
- ask for jobs non manipulated by "cmsunified" (a CMS internal flag needed to trust the workload description);
- end user analysis jobs have been vetoed; they are typically more complex to control for what concerns memory, io and time utilization;
- ask for resizeable jobs, which are explicitly allowed to be run in a range of thread configurations. This has been chosen in view of further manipulating the payloads, as explained later.

The current regular expression used to cherry-pick among CNAF-directed jobs is

```
(regexp("^(?!.*cmsunified).*$",WMAgent_SubtaskName) &&
regexp("wmLHEGS-[^/]*$",WMAgent_SubtaskName) && (WMCore_ResizeJob == True
|| RequestMemory/RequestCPUs < 2000)) ||
(regexp("^(?!.*cmsunified).*$",WMAgent_SubtaskName) &&
regexp("wmLHEGEN-[^/]*$",WMAgent_SubtaskName))
```

The CMS Software Framework (CMSSW) has been in production with multi-threaded processes since 2015, using Intel TBB [19] as threading engine.

The capability to increase the number of threads in a wide range (from 1 to 8 threads are generally used in production, with an higher thread count avoided just not to be too impacted by the Amdahl's law [20]); in the KNL case, this is usable in two ways:
- to decrease the memory/node needs: a KNL noda has < 1.5 GB/core or even < 0.5 GB/thread if hyperthreading is on. Massively multithreaded processes are able to fit even this pressing requirement.
- to decrease process run time. CMS payloads are tuned to execute in 8 hours on standard Xeon-like systems. A KNL core is 3x-4x slower, and hence the jobs would exceed the 24 h slot time in SLURM. An increase in process thread count allows to shorten the run times.

Tests at the project submission have shown that, by increasing 4x the thread count per process (from 8 to 32), timing was similar to Xeon nodes, memory was well fitting the 96 GB. The performance hit caused by the increased multithreading was evaluated as 20%, measured as events per second per node in the two 8 and 32 threads configurations.

Test working points with different settings for *(# of Threads per Process, # of Processes per node)* are shown in Figure 2 (right), together with the impact on throughput (*y*-axis) and memory utilization (*x*-axis).

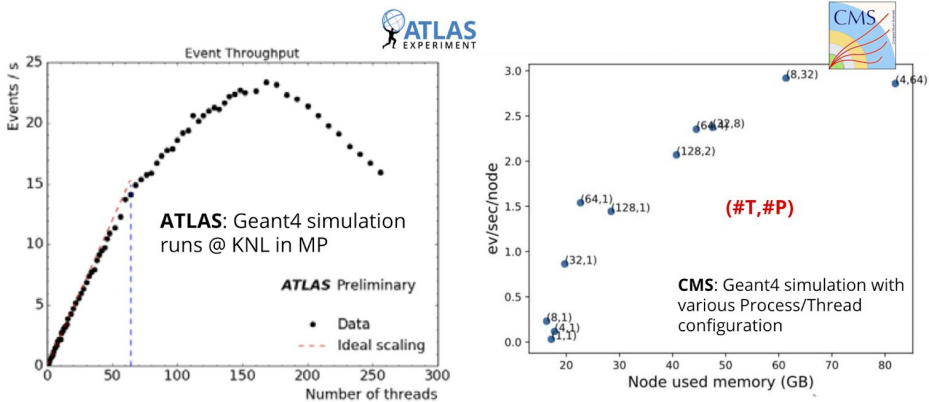

**Fig. 2.** Left: event/s from ATLAS MP simulation on KNL, up to HT4x; Right: throughput and memory utilization (Minimum Bias events) with varying configurations in (*(# of Threads per Process, # of Processes per node)*.

## 5.4 LHCb

In LHCb, Monte Carlo (MC) simulation jobs, which include both event generation and detector simulation, are by far the largest consumers of the experiment's share of WLCG compute resources (they will be over 90% in Run3). The LHCb strategy for exploiting any newly available compute resources is, therefore, to use them for MC simulation.

The integration of new compute resources into the LHCb distributed computing framework, based on DIRAC [21], is generally an easy task when:
- worker nodes (WNs) have outbound network connectivity;
- the LHCb CVMFS endpoints are mounted on the WNs;
- the WN O/S is an SLC6 or CC7 compatible flavor, or Singularity containers are available.

All these three conditions were met on Marconi A2.

The main problem that had to be addressed to submit LHCb jobs on Marconi A2, however, is the fact that this system uses many-core KNL processors with a very limited amount of memory per hardware thread (300 MB if 4 threads are used per physical core, or 600 MB if only 2 threads are used per physical core). This required significant integration efforts, both on the software and on the computing side (more details about this work were given in an LHCb presentation [22] at this conference). Until recently, in fact, all of the LHCb software workflows submitted to distributed compute resources, including MC simulation, only consisted of single-process (SP), single-threaded (ST) applications. These workflows are very inefficient on KNL, because they have a much larger memory footprint (approximately 1 GB per thread), and on the software side it was therefore necessary to re-engineer the LHCb simulation software to use multi-processing (MP), or multi-threading (MT). In addition, since this is the first time MP or MT applications are used by LHCb on distributed resources, on the computing side the LHCb DIRAC framework had to be adapted to be able to manage the scheduling of jobs using more than one hardware thread.

More in detail, the software work targeting the Marconi A2 timescales focused on the test and commissioning of a multi-processing based version (GaussMP) of the LHCb simulation framework (Gauss). This leveraged on a multi-processing version (GaudiMP) of the LHCb event processing framework Gaudi, already existing but never used in production. Extensive testing and bug-fixing was therefore required, including a validation of results requiring identical results in the SP and MP versions of Gauss when simulating the same set of events starting from the same random number seeds. Standalone performance tests on the Marconi A2 KNL nodes showed that the overall throughput of events processed per unit time on one node is highest when 8 GaussMP jobs are executed in parallel, each using 17 processes. This corresponds to using 2 hardware threads per KNL core (i.e. 136 in total): no increase in throughput was observed when trying to use 4 threads per KNL core (i.e. 272 in total), and many failures were actually observed in this configuration. If SP Gauss is used, the maximum throughput was achieved when using only 85 processes (i.e. 85 hardware threads). The best GaussMP throughput is only moderately higher (15%) than that achieved using SP Gauss, because the forking strategy used has not yet been optimized. Looking forward, however, LHCb software efforts in the simulation area will focus not on GaussMP but rather on a MT based solution, Gaussino [23], which will ensure a much lower memory footprint per thread. Gaussino was not ready in time for production use on Marconi A2, but has significantly progressed since CHEP2019 and LHCb plans to test it on the remaining time available on Marconi A2 in 2020.

On the computing side, the main challenge LHCb had to address was that each job slot provided by the HTCondor CE represents a whole KNL node from the A2 partition, with 68 physical cores and up to 4 hardware threads per core. The decision on how to subdivide this allocation between several applications is delegated to each experiment's framework, specifically to DIRAC in the case of LHCb. Rather than implementing a quick ad-hoc solution for Marconi A2, this was addressed in DIRAC by implementing a very generic mechanism for managing "fat nodes", which keeps into account the possibility that a variety of SP/ST jobs using only one hardware thread, and of MP or MT jobs using more than one thread, may coexist simultaneously on the same node. This new functionality has been successfully tested in a dedicated certification environment, but at the time of writing it has not yet been deployed in production for LHCb, because of the timescales involved in the experiment's software release process. This is the reason why no results of production use of Marconi A2 by LHCb are shown in section 6 of this paper. LHCb plans to submit its first production of MC simulation jobs on Marconi A2 as soon as the new LHCb DIRAC is released, using software workloads based on GaussMP, or possibly Gaussino if available.

## 6 Results

At the time of writing, only the CMS experiment has been able to execute workflows in production. Results are shown from the first period (mid Dec19 - mid Jan20), and have been limited to 150 simultaneous nodes; in the same period, in order to avoid stressing the storage system, the number of used cores has been limited to 64, without hyperthreading[1]. Hence, a total of 150x64 = 9600 threads have been accessible.

Figure 3 shows that the system has been able to sustain processing at 9600 threads for (many) days; the moments without jobs were either due to site problems, or to scarcity of compatible payloads in the CMS queues.

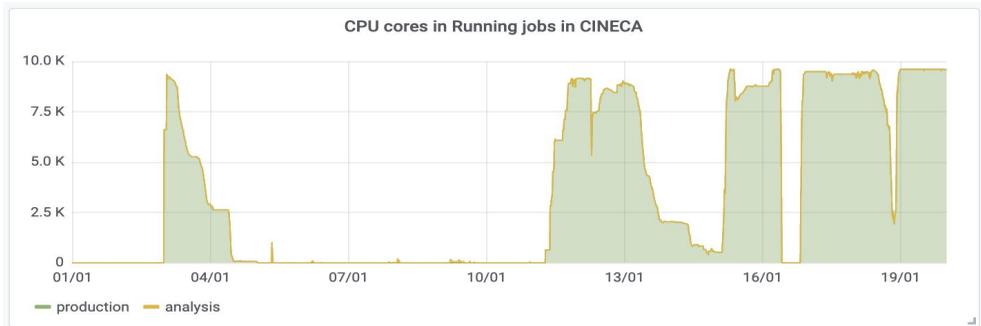

**Fig. 3.** Number of cores running for CMS in the period Dec19-Jan20.

Figure 4 shows in a pie chart the success rate of such jobs from CMS monitoring.

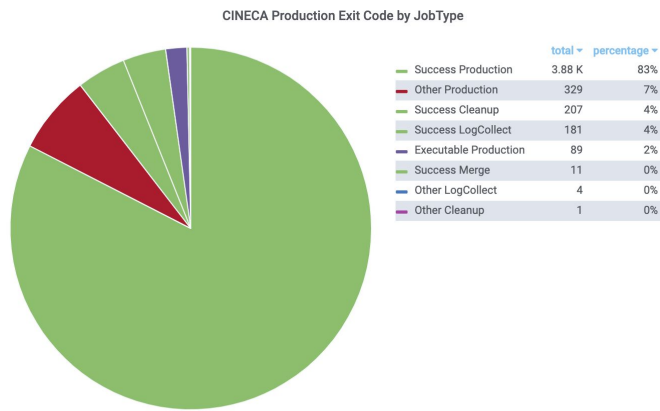

**Fig. 4.** Exit code for CMS jobs running at CINECA.

The overall result (92% of successful jobs) is compatible with what CMS sees at more standard sites. In all the periods, thanks to the precise cherry picking of jobs with specific needs, CINECA general network has not suffered from the additional traffic.
Figure 5 shows a view of the CNAF facility from the point of view of CMS. Since CINECA's resources have been added to standard ones, CMS sees CNAF as a processing site able to serve > 17k threads (~7k at CNAF and 10k at CINECA).

---

[1] In a realistic production, it was seen that the help of multithreading was much lower than in the initial test case.

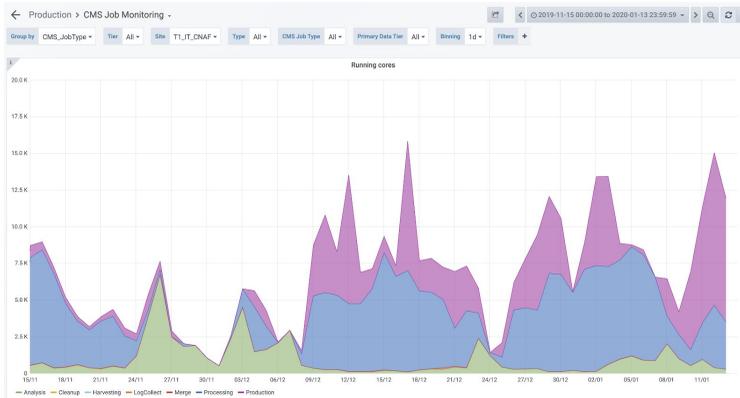

**Fig. 5.** CMS view of CNAF facility. Since the integration of CINECA has happened as an extension of CNAF, the total number of slots is reported. The violet peaks, with the total number of cores exceeding 15k, are mostly coming from CINECA slots.

## 7 Future directions and conclusions

At the time of writing, the Marconi A2 system is off. Our project has currently used 20% of the grant hours, and a restart of the operations is expected around April 2020.
It should be noted that, in recent months, most of Marconi A2 has been switched off, as a significant fraction of the compute capacity has been moved from KNLs in Marconi A2 to GPUs in a new Marconi100 partition. This is a move which could have had a large impact on our work if all KNL nodes had been switched off, as none of the software workflows described in this paper by any of the four LHC experiments has yet been ported to GPUs and/or validated for production use on a GPU-based computing system. Software development work on GPUs is currently very active in all experiments and in common software projects, but it may still take months, or years, before large-scale use of GPU compute resources by the LHC experiments is possible. This should also be taken into account in the planning of future facilities.
Our plans are to conclude the grant utilization on Marconi A2's remaining KNL nodes, putting in production also the other LHC experiments, and to learn more on aspects like data caching at the CNAF site, also in collaboration with the efforts from the ESCAPE EU Project and the definition of a data lake testbed.

The collaboration with CINECA has been very fruitful for the LHC experiments, since it has driven a number of R&D needed in order to optimally utilize such a non standard machine. At the same time, we think it will be even more essential in the design phase of the next CINECA HPC, Leonardo, which will have LHC Experiments between its target use cases from day one.

The present work is a continuation of a series of integration efforts, as reported in [24], [25], [26].

This work has been partially supported by the ESCAPE EU project [27], G.A. 824064.